\documentclass{article}
\usepackage{spconf,amsmath,graphicx}
\usepackage{xcolor}

\usepackage{enumitem}
\setlist{nosep, leftmargin=14pt}

\usepackage{mwe} 

\usepackage{multirow}

\title{Learning Population-level Shape Statistics and Anatomy Segmentation From Images: A Joint Deep Learning Model}
\name{Author(s) Name(s)}
\address{Author Affiliation(s)}
\name{Wenzheng Tao $^{\star \dagger}$ \qquad Riddhish Bhalodia $^{\star \dagger}$ \qquad Shireen Elhabian $^{\star \dagger}$ \qquad}
\address{
$^{\star}$
Scientific Computing and Imaging Institute, University of Utah, Salt Lake City, Utah-84112, USA \\
$^{\dagger}$ School of Computing, University of Utah, Salt Lake City, Utah-84112, USA}
\begin{document}
%
\maketitle
\begin{abstract}
Statistical shape modeling is an essential tool for the quantitative analysis of anatomical populations. Point distribution models (PDMs) represent the anatomical surface via a dense set of correspondences, an intuitive and easy-to-use shape representation for subsequent applications. 
These correspondences are exhibited in two coordinate spaces: the local coordinates describing the geometrical features of each individual anatomical surface and the world coordinates representing the population-level statistical shape information after removing global alignment differences across samples in the given cohort. 
We propose a deep-learning-based framework that simultaneously learns these two coordinate spaces directly from the volumetric images. The proposed joint model serves a dual purpose; the world correspondences can directly be used for shape analysis applications, circumventing the heavy pre-processing and segmentation involved in traditional PDM models. Additionally, the local correspondences can be used for anatomy segmentation. We demonstrate the efficacy of this joint model for both shape modeling applications on two datasets and its utility in inferring the anatomical surface.
\end{abstract}
\begin{keywords}
Deep learning, statistical shape analysis, anatomy segmentation
\end{keywords}
\vspace{-0.20in}

\section{Introduction}
\vspace{-0.15in}
Statistical shape modeling (SSM) represents the study of shape characteristics given a sample population. It finds a wide range of applications, such as in pathology characterization \cite{harris2013cam}, growth modeling \cite{datar2009particle} and implant design \cite{goparaju2018evaluation}. 
Shape representation is core of SSM, where each anatomy is parsed into a population consistent quantitative form. Shapes can be represented explicitly, via an ordered set of points (aka point distribution models or PDMs) that are placed in corresponding locations across all instances \cite{cates2007shape,RTW:Gre91}, or implicitly via  deformation field to an atlas \cite{beg2005computing}. 

Deformation fields used for shape analysis (e.g., \cite{durrleman2014morphometry}) are hard to interpret and usually require a notion of an atlas \cite{joshi2012diffeomorphic}, that can be hard to obtain. On the other hand, PDMs are intuitive and easy to use, and several algorithms exist to generate PDM automatically. One example is the entropy-based methods (e.g., \cite{cates2007shape, davies2002MDL}) that perform a non-linear optimization to manipulate the points positions on the given surfaces using an objective function that enforces model statistical compactness. Another example is the approaches that rely on pairwise mapping to a geometric primitive, such as SHPARM-PDM \cite{styner2006spharm}, which maps shapes to a unit sphere using a spherical harmonic based shape representation. However, all these algorithms require segmented anatomies that hinge on the availability of manpower and domain-expertise with the specific anatomy to perform such time-consuming task \cite{bhalodia2018deepssm}. Additionally, these PDM algorithms require a pre-processing pipeline involving re-sampling, rigid alignment, and smoothing to factor out any non-anatomical artifacts before shape modeling, which is expensive in terms of computation and human resources. 
%
%
Surface correspondences in a PDM entail two coordinate spaces \cite{cates2017shapeworks}; a configuration space where such correspondences are represented in the \textit{local} coordinates of each surface, and a shape space where global alignment differences are factored out such that all shapes share the same \textit{world} coordinate system for population-level shape analysis. 


Recently, neural networks have been applied for shape representation. Unsupervised registration (e.g., \cite{balakrishnan2018unsupervised}) showcase the computational efficacy of neural networks for registration and deformation based shape analysis. Similarly, for PDM, some recent works (e.g., \cite{bhalodia2018deepssm, adams2020uncertain, bhalodia2021deepssm}, have proposed a deep-learning-based approach to circumvent the resource-heavy computation in the PDM model generation. These methods use an existing PDM to train a deep network that captures the mapping from volumetric images to shape representation; which mitigating segmentation and pre-processing during inference on a new image and facilitates a fully end-to-end automatic pipeline for shape analysis. The models are geared towards SSM and are trained to predict the world coordinates of the PDM. However, when predicted directly from a 3D image, the local coordinates of the PDM also have utility. They can be used for anatomy segmentation/coarse estimation of the anatomical surface to provide a shape prior for segmentation algorithms. 

In this paper, we propose a novel extension of the deep learning framework in \cite{bhalodia2018deepssm} to predict both the local and world coordinates of the PDM jointly. This joint estimation reduces the training by half, also the joint learning of world and local coordinates regularizes both the population shape representation as well as subject-specific anatomy representation. Learning the local coordinates is improved by world coordinates, as the world coordinates act as a population level shape prior. On other hand the world coordinate representation is regularized by an learning of the anatomical surface utilizing local coordinates. 
We evaluate the proposed model's efficacy in shape modeling downstream applications and anatomy segmentation with baseline models trained individually on the world and local coordinates, respectively.

\vspace{-0.1in}

\section{Methods}
\vspace{-0.15in}
Given a set of images and segmentations, we use \emph{ShapeWorks} \cite{cates2017shapeworks} to obtain the PDM model of the given anatomical cohort.  
We obtain the \emph{world} and \emph{local} correspondences, denoted by $P_i$ and $Q_i$, respectively, of the segmentation of the $i$-th image $I_i$ using a transformation $T_i: P_i \rightarrow Q_i$ that captures rotation, translation, and scale using generalized Procrustes alignment \cite{gower1975generalized}. 
The proposed network simultaneously learns the mapping from volumetric images  $\{I_i$\} to the world particles $\{P_i\}$ and the corresponding transformations $\{T_i$\}.

\textbf{Data Augmentation: }Typically, shape modeling tasks in medical applications do not have large samples sizes to train data-hungry deep networks. We utilize data-augmentation similar to one proposed in \cite{bhalodia2021deepssm}.
Specifically, we perform Principal Component Analysis (PCA) on $ \{P_i\} $ to obtain low-dimensional PCA scores $ \{Z_i \in R^{L \times 1} \} $, where $L$ is one less than number of samples, as we utilize the full PCA space. We use kernel density estimation (KDE) similar to \cite{adams2020uncertain, bhalodia2021deepssm} to estimate the distribution density in the PCA subspace. 
The KDE allows us to define a density from which we can sample new PCA scores $\{Z_j\}$ and reconstruct new correspondences $\{P_j\}$ using PCA reconstruction operator.
%
%
Our augmentation process still needs the corresponding transformation and image for the augmented shapes (i.e., world points $\{P_j\}$). Hence, we also augment transformations
. We
decompose each 
transformation as a 7-dimensional vector that includes a rotation vector, a translation vector, and a global scaling factor. 
After augmentation, unit quaternions are used for rotations to avoid rotational ambiguity. We model the original transformations using a Normal distribution and sample every new $ T_j $ as $T_j \sim \mathcal{N} ( \mu_T, \Sigma_T )$ where $ \mu_T \in R^{7 \times 1} $ and $ \Sigma_T \in R^{7 \times 7} $ are the sample mean and diagonal covariance matrix. Local points of the $j-$th augmented sample $Q_j$ are then obtained by applying $ T_j $ to $ P_j $. For its corresponding image $ I_j $, the Thin plate spline (TPS) warp from $ Q_i $ to $ Q_j $ is used to deform $ I_i $, where $Q_i$ is the shape whose kernel was used to sample $Q_j$. 



\begin{figure}[!h]
  \centering
  \includegraphics[width=0.45\textwidth]{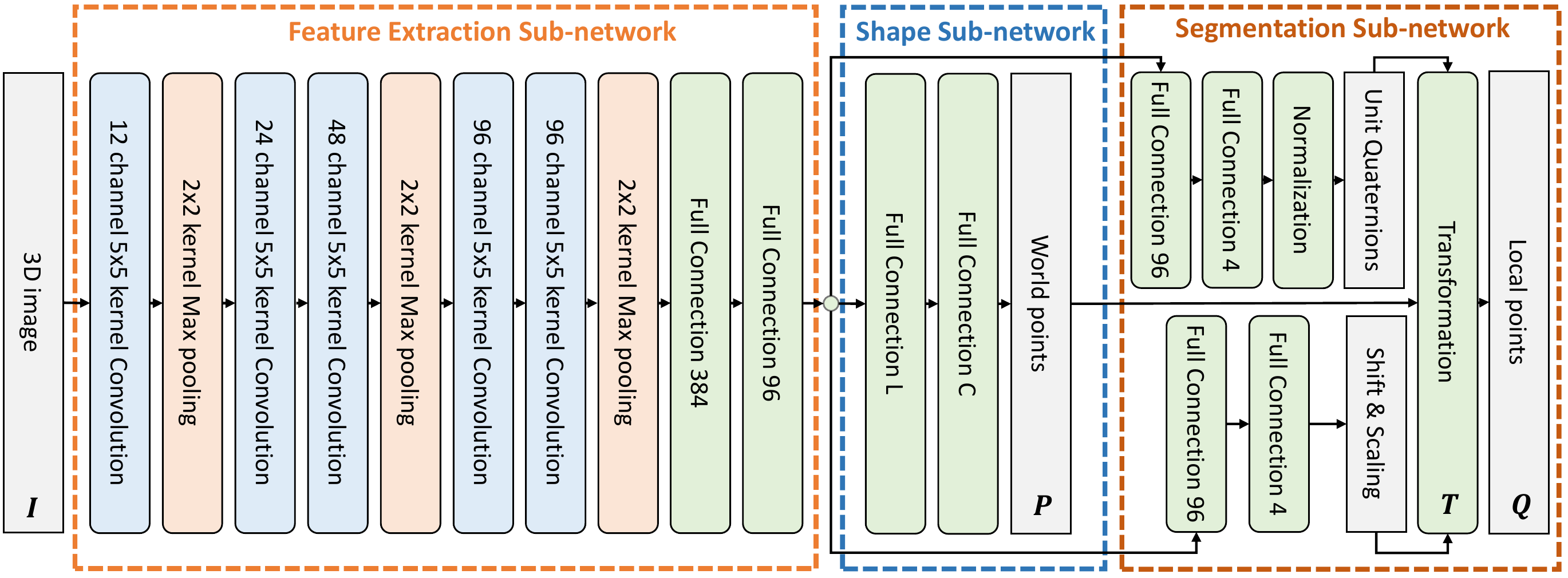}
 \vspace{-0.5cm}
\caption{\textbf{Network architecture.} It is composed of 3 sub-networks, backbone, shape and segmentation. }
\vspace{-0.4cm}
\label{fig:net_arch}
\end{figure}


\textbf{Network Architecture and Training: } We use a deep neural network to learn the population level shape representation via world correspondences and the subject specific anatomical representation or segmentation via local correspondences, 
as shown in Fig. \ref{fig:net_arch}. 
The backbone is the feature-extraction sub-network, it has 5 convolution, 3 max pooling, and 2 full connected layers. It first increases the channels to 12 with a size factor (1/8) and then extracts a low dimensional feature vector.  The extracted features  are utilized for two different sub-networks, described as follows:
\begin{enumerate}
    \item \textbf{Shape sub-network: } estimates world correspondences that represent the population-level shape information. It is a set of 2 fully-connected layers that regress to the PCA scores $\hat{Z}_i$. The PCA score is passed through another linear fully-connected layer with the weights fixed as the PCA bases (i.e., eigen vectors) and the bias as the mean shape, and hence, this layer performs PCA reconstruction giving us an estimate of the world coordinates $\hat{P}_i$.
    \item \textbf{Segmentation  sub-network: }  estimates the transformations $\hat{T}_i$ and then using the predicted world correspondences, the local correspondences $\hat{Q}_i$ are estimated, which describe the anatomical surface. First it learns the transformation with 4 fully connected layers to regress the transformation parameters from the learned features of the backbone sub-network. There are 8 transformation parameters $ \hat{T}_i $ to predict, 4 for unit quaternions, 3 for translations and 1 for scaling factors. The quaternions are normalized and the real component is constrained to be non-negative to avoid ambiguity.
\end{enumerate}



The proposed network is trained using a coarse-to-fine training strategy. 
For the coarse training, we train the model by a relative mean squared error loss $ \mathcal{L}_1 $ in Eq. (\ref{eq:loss})(N=training size), $ \oslash $ denotes Hadamard division, $ \sigma_T \in R^{8 \times 1} $ and $ \sigma_Z \in R^{L \times 1} $ are the empirical standard deviations of transformations and PCA scores, respectively. $ \bar{T_i} $, $ \bar{Z_i} $, $ \bar{Q_i} $ are the empirical means. 
As the transformation loss $ \mathcal{L}_T $ and the PCA scores loss $ \mathcal{L}_Z $ are relative losses, they are commensurate during training.
For refinement we start with the coarse model but we utilize a single loss of supervising over the local coordinates, given by $\mathcal{L}_2$ in Eq. \ref{eq:loss}. This loss will continue to update the predictions for both the world coordinates and transformations.
We employ this coarse-to-fine training methodology as we observed that direct supervision on the local correspondences leads the network to underfit.

\vspace{-0.70cm}
\begin{align}
    \mathcal{L}_1 & = f(T, \sigma_T) + f(Z, \sigma_Z),
    \mathcal{L}_2 = f(Q, \mathbf{1})
    \label{eq:loss}\\
    f(X, \sigma_X) & = \frac{ \frac{1}{N} \sum_{i=1}^N || (X_i - \hat{X_i} ) \oslash \sigma_X ||_2^2 }{ \frac{1}{N} \sum_{i=1}^N || (X_i - \bar{X_i} ) \oslash \sigma_X ||_2^2 }
\vspace{-0.4cm}
\end{align}

\vspace{-0.2in}

\section{Results}
\vspace{-0.2in}
This section evaluates the proposed method on two clinical datasets. We compare it with two baseline methods denoted by $ \mathrm{DeepSSM}_{W}$ and $ \mathrm{DeepSSM}_{L}$, respectively. These two baselines are the same DeepSSM model proposed in \cite{bhalodia2018deepssm} trained separately on different input correspondences. The $ \mathrm{DeepSSM}_{W} $ only learns the statistics of shape, i.e., the world correspondences. The $ \mathrm{DeepSSM}_{L} $ learns the anatomical surface, i.e., the local correspondences. Each of these baselines are trained by supervising over world and local coordinates, respectively. By comparing to these baselines, we showcase that the proposed method that jointly learns world and local coordinates performs better or comparatively to the networks trained, separately. 
For each dataset, we evaluate the performance of each model by comparing the predicted correspondences to their ground truths and computed the root mean squared error (RMSE) averaged over each dimension.
%
We also evaluate how well the world coordinates from the proposed model and the baseline perform for downstream shape modeling applications. Similarly, we also evaluate dice scores between the ground truth and meshes discovered from predicted local correspondences; this gives us a measure of how well each model discovers the anatomical surface.



\begin{table}[]
\centering
\begin{tabular}{|l|l|l|l|l|}
\cline{1-5}
    \multicolumn{1}{|l|}{ \textbf{Data set}} & \textbf{Metric} & $ \mathrm{DeepSSM}_{W} $ & $ \mathrm{DeepSSM}_{L} $ & Ours \\ \hline
    \multirow{3}{3em}{\textbf{Cranium}} 
    & \textbf{World} & 0.81 & NA & 0.65 \\ 
    & \textbf{Local} & NA & 1.71 & 1.46 \\ 
    & \textbf{Dice} & NA & 0.97 & 0.97 \\ 
    \hline
    \multirow{3}{3em}{\textbf{Femur}} 
    & \textbf{World} & 0.51 & NA & 0.55 \\ 
    & \textbf{Local} & NA & 0.70 & 1.11 \\ 
    & \textbf{Dice} & NA & 0.95 & 0.93 \\ 
    \hline    
\end{tabular}
\caption{ \textbf{Shape and surface learning accuracy}. World/local particle RMSE is averaged by x,y and z coordinates. Dice coefficient is using mean mesh with local particle predictions. }
\vspace{-0.2in}
\label{tab: score_table}
\end{table}


\noindent\textbf{Metopic Craniosynostosis}

Metopic Craniosynostosis is a morphological abnormality 
that occurs due to early fusion of the metopic suture. Morphological symptoms include a triangular head shape called trigonocephaly. Head deformity evaluation 
is used for diagnosis which plays a key role in treatment planning. We had 120 head CT scans with their cranial segmentation of patients aged between 5-15 months. 28 of these patients were diagnosed with metopic craniosynostosis; while the remaining 92 were normal. Using stratified random sampling, we selected 105 samples for training and 15 for testing. \emph{ShapeWorks} \cite{cates2017shapeworks} package was used to generate the initial PDM with 2048 particles on each shape. We downsampled to $125 \times 91 \times 118$ voxels with isotropic 2.4 mm voxel spacing to fit GPU memory. 
10\% of all training images (including 5000 samples we augmented) form the validation set. 

\begin{figure}[!h]
  \centering
  \centerline{\includegraphics[width=0.45\textwidth]{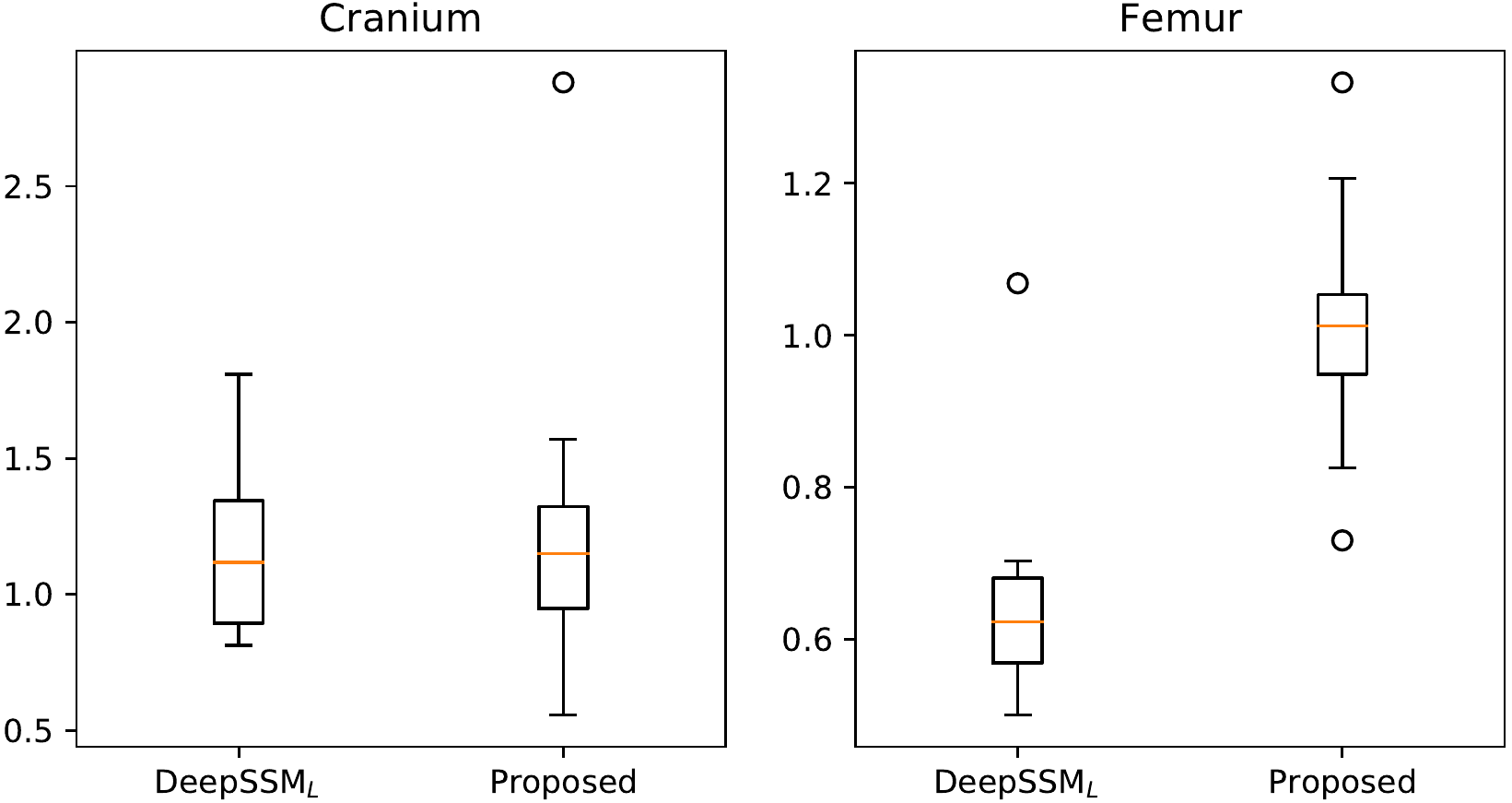}}
\caption{\textbf{Surface-to-surface distance box-plots}. The surface-to-surface distance is computed between ground truth mesh and mesh obtained from predictions across all test samples. The y-axis is the surface-to-surface distance in millimeters. }
\vspace{-0.4cm}
\label{fig:boxplot}
\end{figure}

\begin{figure*}
    \centering
    \includegraphics[width=0.8\textwidth]{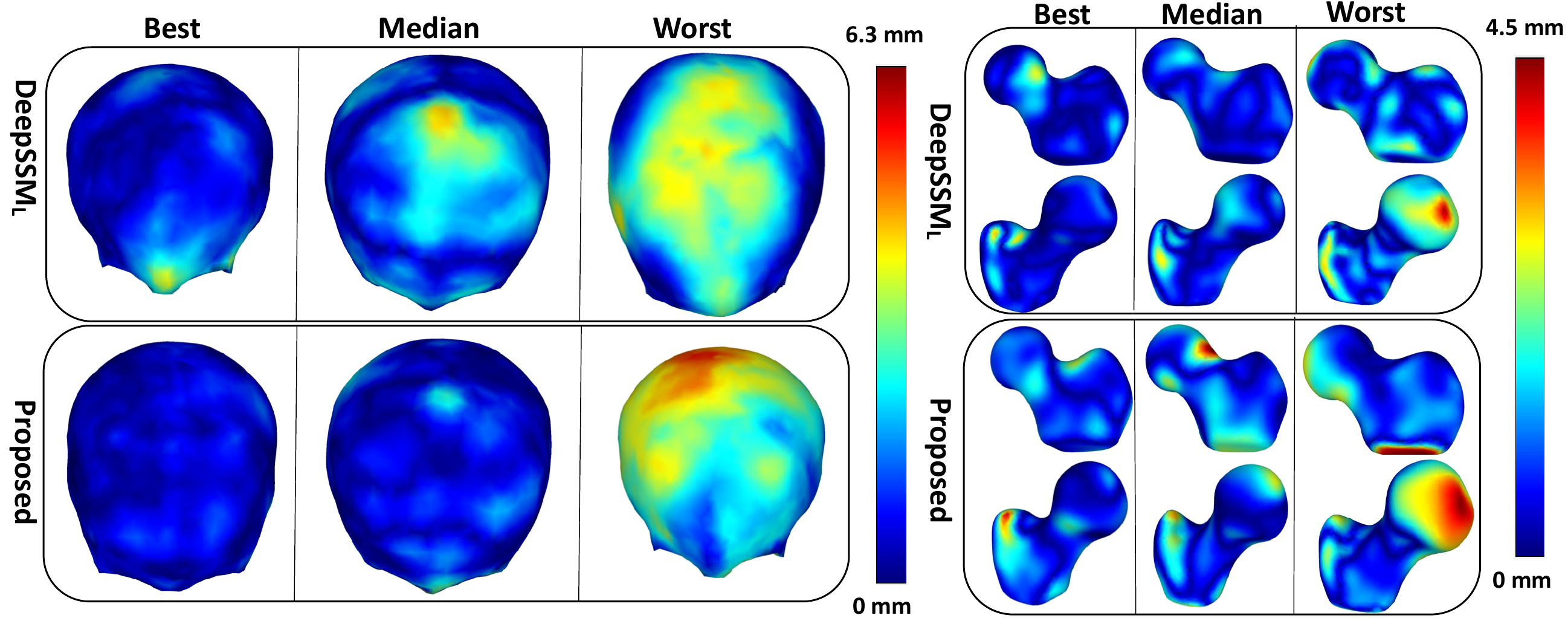}
	\caption{\textbf{Surface-to-surface distance}. It is visualized on the ground truth surfaces,  samples chosen separately for each method.}
\vspace{-0.4cm}
    \label{fig:surface_dist}
    
\end{figure*}

We evaluated the baselines and the proposed model and the RMSE scores and Dice scores are shown in  Table \ref{tab: score_table} and surface-to-surface distance in \ref{fig:boxplot}. We see that the proposed joint model performs comparably or better to the baselines. Specifically, for local correspondence estimation the proposed model outperforms the $\textrm{DeepSSM}_L$ baseline due to the shape prior like regularization that joint training provides and reduces error in high frequency variations. 
The visualization of the surface-to-surface distance for the test scans with worst, median and best performance are shown in Figure \ref{fig:surface_dist}.
 

To evaluate the performance of the world correspondences in downstream shape modeling tasks, we performed severity quantification of the metopic scans. 
For severity, we referred to \cite{tao2020unsupervised} where they proposed an unsupervised shape deformity measure for metopic craniosynostosis. Following their methodology, we modeled 72 control scans with  Probabilistic Principal Component Analysis (with 95\% variance subspace). Using this model for the column and null space, we computed the Mahalonobis distance using the learned covariance matrix, and this distance is an unsupervised measure of the severity. 
To evaluate the quantified severity, we utilized ratings for the 48 scans from 36 craniofacial experts in a 5 point Likert scale aggregated using Latent trait theory \cite{uebersax1993latent}. We computed correlation of the estimated severity with this aggregate obtained from experts. We also computed the area under curve (AUC) using the binary diagnosis for the estimated severity. Results are compiled in Table. \ref{tab:cranio_table}. We see that the proposed method performs comparatively with world baseline and Shapeworks, hence, showcasing that the proposed model can capture the statistical shape information via world correspondences well to perform any analysis application.
\begin{figure}[htb]
    \centering
    \includegraphics[width=0.4\textwidth]{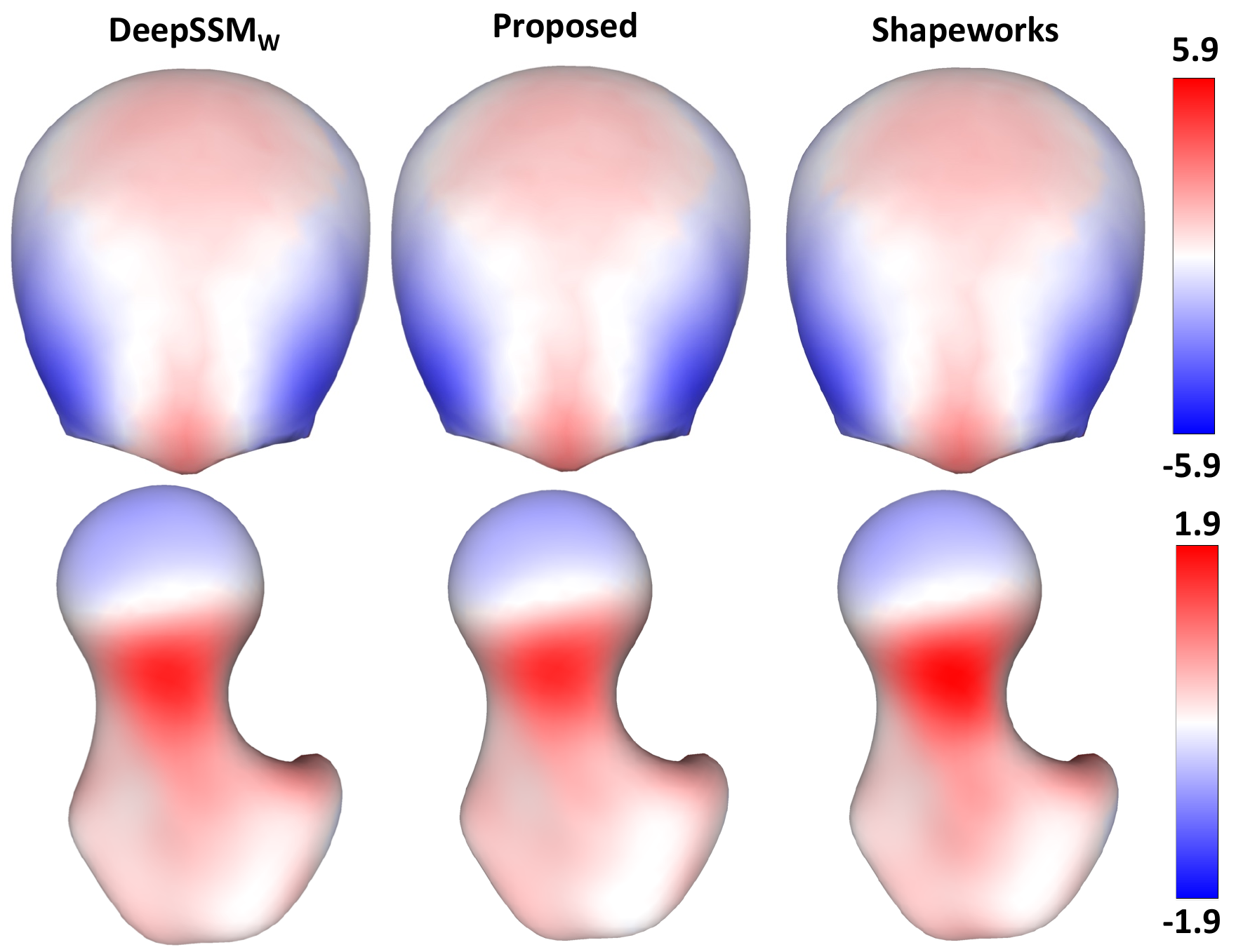}
    \caption{\textbf{Group difference}.  Red denotes outwards deformation and blue denote the opposite.}
    \vspace{-0.4cm}
    \label{fig:group_diff}
    
\end{figure}

To characterize the metopic craniosynostosis deformity, we computed \emph{group differences} \cite{harris2013cam}, taking difference between means of normals and pathological shapes. We showcase this group difference as a heatmap on the normal mean mesh in Fig. \ref{fig:group_diff}. All the three methods recognize the proper trigonocephaly deformity for metopic craniosynostosis, indicating that proposed methods captures pathological variation well. 



\begin{table}[]
\centering
\begin{tabular}{l|l|l|l|}
\cline{1-4}
    \multicolumn{1}{|l|}{ \textbf{ Metric }}  & $ \mathrm{DeepSSM}_{W} $ & Ours & ShapeWorks \cite{cates2017shapeworks} \\ \hline
    \multicolumn{1}{|l|}{\textbf{PearsonR}}  & 0.78  & 0.84 & 0.85 \\ \hline
    \multicolumn{1}{|l|}{\textbf{SpearmanR}}  & 0.69  & 0.79 & 0.83 \\ \hline
    \multicolumn{1}{|l|}{\textbf{AUC}}  & 0.81 & 0.89 & 0.93 \\ \hline
\end{tabular}
\caption{\textbf{Evaluation of estimated severity:}  Correlation and AUC of the severity measure with respect to the expert ratings and diagnosis. Higher numbers are better.}
\vspace{-0.4cm}
\label{tab:cranio_table}
\end{table}


\noindent\textbf{Cam-type Femoroacetabular Impingement }


Cam-type femoroacetabular impingement (cam-FAI) is characterized by an abnormal 
femoral head
shape
. Local characterization of the deformity by comparing it with normal shape population can provide valuable guidance to clinicians for treatment planning. We had 49 samples for training (42 controls, 7 cam-FAI), 9 for  testing (7 controls,2 cam-FAI). We used \emph{ShapeWorks} \cite{cates2017shapeworks} to get 1024 correspondences as initial PDM. We downsampled images to $117 \times 82 \times 105$ voxels with isotropic 1.2 mm voxel spacing to fit GPU memory. 
Due to little transformation variability in the training data, we manually specified diagonals of $ \Sigma_T $ as 
$ [1e^{-4}, 1e^{-4}, 1e^{-4}, 1, 1, 1, 1e^{-3}] $
. 10\% of all training images (including 5000 augmented) were left for validation.

Similarly to metopic dataset, Table. \ref{tab: score_table} shows the RMSE and dice score performances , and the Figures \ref{fig:boxplot} and \ref{fig:surface_dist} showcase the surface-to-surface distances. 
All the results showcase comparable performance of the proposed joint model with the individually trained baselines. 
Fig. \ref{fig:surface_dist} reveals that the majority of 
its errors
are coming from inconsistencies in defining the cutting plane to crop the proximal femur. 
Specifically if we take the worst case femur example from Figure \ref{fig:surface_dist}. This scan estimates a large scaling factor to account for the extremely large femoral head, which resulted in predicting a lower cropping plane.
The scan in question has an unnaturally cropped plane, a variation which is not captured by the original PDM model. Since the proposed method learns segmentations that are regularized by the shape statistics in world correspondences, it ignores this variation. This regularization via shape is usually advantageous, and these test scans represent an exception. In contrast $\mathrm{DeepSSM}_L$ without any regularization is able to capture these exceptions better, but suffers in making smooth predictions at local anatomies of interest.

To evaluate world correspondences in downstream application, we characterized the cam-FAI deformity via \emph{group-differences} \cite{harris2013cam}. This group difference is overlayed on top of a mesh as a heat map.  Fig. \ref{fig:group_diff} showcases that the proposed method discovers a similar pathological pattern as discovered by Shapeworks PDM, and the 
$ \mathrm{DeepSSM}_{W} $ baseline. 



\vspace{-0.1in}

\section{Conclusions}
\vspace{-0.2in}
We proposed a supervised deep learning based method that is jointly trained to learn (i) population shape representation in form of \emph{world} correspondence that removes location, rigid rotation and scaling, (ii) subject-specific anatomical representation via \emph{local} correspondences in the configuration space which captures the surface/segmentation. The joint training of shape and segmentation provide a bi-directional regularization. Segmentation is regularized by learning the world coordinates as they act like a statistical shape prior. Similarly, the shape representation is regularized by local correspondences as it informs surface. We compare the proposed model to DeepSSM \cite{bhalodia2018deepssm}, a state-of-the-art deep learning model to learn correspondences from images, with two separate models trained on world and local correspondences. We find that the qualitative and quantitative performance of the proposed model is better or comparable with these baselines on two different datasets. We also evaluate dice scores and surface-to-surface distances for segmentation performance, and downstream applications for shape analysis using world coordinates for shape modeling performance. In these comparisons, the proposed model performs better/comparably to baselines.

\vspace{-0.1in}

\section{Compliance with Ethical Standards}

This study was performed in line with the principles of the Declaration of Helsinki. 
\vspace{-2pt}

\section{Acknowledgments}
This work was supported by the National Institutes of Health under grant numbers of R21-EB026061, NIBIB-U24EB029011, NIAMS-R01AR076120, NHLBI-R01HL135568, NIBIB-R01EB016701.
The content is solely the responsibility of the authors and does not necessarily represent the official views of the National Institutes of Health.

\bibliographystyle{IEEEbib}
\bibliography{strings,refs}

\end{document}